\pdfoutput=1

\documentclass{llncs}
\usepackage[T1]{fontenc}    
\usepackage{hyperref}       
\usepackage{url}            
\usepackage{booktabs}       
\usepackage{amsfonts}       
\usepackage{amsmath}
\usepackage{nicefrac}       
\usepackage{microtype}      
\usepackage{xcolor}         
\usepackage{xspace}
\usepackage{subcaption}
\usepackage{comment}
\usepackage{tikz}
\usepackage{cleveref}
\usepackage{multirow}

\usepackage[n, advantage , operators ,
    sets ,
    adversary ,
    landau ,
    probability ,
    notions ,
    logic ,
    ff ,
    mm,
    primitives ,
    events ,
    complexity ,
    asymptotics ,
    keys
]{cryptocode}

\newcommand{\R}{\ensuremath{\mathbb{R}}\xspace}

\newcommand*{\numero}{n\kern-.1em \raise.7ex\vbox{\hbox{\tiny \ensuremath{\circ}}\kern.5pt}}

\hypersetup{colorlinks=true,
            urlcolor=blue,
            linkcolor=black,
            citecolor=blue,
            breaklinks=true}

\title{CryptoTL: Private, Efficient and Secure Transfer Learning}

\author{%
  Roman Walch\inst{1,2}\and
  Samuel Sousa\inst{1,2}\and
  Lukas Helminger\inst{1,2}\and
  Stefanie Lindstaedt\inst{1,2}\and
  Christian Rechberger\inst{1}\and
  Andreas Tr\"ugler\inst{1,2}
}%
\institute{%
  Graz University of Technology (Austria)\and
  Know-Center GmbH (Austria)\newline
  roman.walch@iaik.tugraz.at\newline
  ssousa@know-center.at\newline
  lukas.helminger@iaik.tugraz.at\newline
  slind@know-center.at \newline
  christian.rechberger@tugraz.at\newline
  atruegler@know-center.at
}%

\begin{document}

\maketitle

\begin{abstract}
   Big data has been a pervasive catchphrase in recent years, but dealing with data scarcity has become a crucial question for many real-world deep learning (DL) applications.
   A popular methodology to efficiently enable the training of DL models to perform tasks in scenarios with low availability of data is transfer learning (TL). TL allows to transfer knowledge from a general domain to a specific target one. However, such a knowledge transfer may put privacy at risk when it comes to sensitive or private data.
   With CryptoTL we introduce a solution to this problem, and show for the first time a cryptographic privacy-preserving TL approach based on homomorphic encryption that is efficient and feasible for real-world use cases. We achieve this by carefully designing the framework such that training is always done in plain while still profiting from the privacy gained by homomorphic encryption. To demonstrate the efficiency of our framework, we instantiate it with the popular CKKS HE scheme and apply CryptoTL to classification tasks with small datasets and show the applicability of our approach for sentiment analysis and spam detection. Additionally, we highlight how our approach can be combined with differential privacy to further increase the security guarantees.
   Our extensive benchmarks show that using CryptoTL leads to high accuracy while still having practical fine-tuning and classification runtimes despite using homomorphic encryption. Concretely, one forward-pass through the encrypted layers of our setup takes roughly $1\,s$ on a notebook CPU.
\end{abstract}

\section{Introduction}\label{sec:introduction}

Deep learning (DL) architectures have been prompting a revolution across many fields, redefining state-of-the-art results, and influencing the lives of many people worldwide with real-time systems for automatic decision-making.
However, some tasks do not benefit from DL's high performance due to data scarcity, which hinders the model training from scratch.
To overcome this problem, transfer learning (TL)~\cite{pan2009survey} has appeared as a methodology for enhancing the performance of DL models for target domain tasks by transferring knowledge from different but related source domains~\cite{ben2010theory,zhuang2020comprehensive}.
TL alleviates the problem of insufficient training data by relaxing the hypothesis that training and test data should be independent and identically distributed~\cite{tan2018survey}.
Therefore, models do not need to be retrained from scratch when used for tasks that differ from those they were originally developed for.

Despite the gains in model performance, the knowledge transfer may be an issue for data privacy.
For instance, model inversion attacks can recover original training samples, even without initial clues about the data or model parameters in learning settings with distributed parties~\cite{heetal2009modelinversion,abadi2016deep}.
Consequently, TL methods that are trained on private or personal data must implement countermeasures against these privacy attacks like differential privacy (DP)~\cite{dwork2008differential}.
Similar privacy challenges also arise in Machine-Learning-as-a-Service (MLaaS), where a server stores a private model trained for a task in a specific domain.
The server monetizes this model by letting clients query results for their input data.
In this sense, maintaining the privacy of the model is a business requirement.

Privacy, however, is a requirement not limited to corporate models. Clients of MlaaS may also refuse to send their potential private input data to servers without protection. In fact, the right to privacy has become a global concern shown by regulations, such as the European Union's General Data Protection Regulation (GDPR)~\cite{EUdataregulations2018} and the California Consumer Privacy Act (CCPA)~\cite{BAIK2020101431}. Those regulations usually demand safeguarding measures to avoid information breaches that cause hazardous situations, such as revealing personal data, damaging the reputation of individuals and companies, or the application of penalties and fines.

\subsection{Privacy-Preserving Primitives}\label{sec:HE}

Privacy-preserving cryptographic protocols and primitives, like homomorphic encryption (HE)~\cite{Rivest1978,DBLP:conf/stoc/Gentry09} and differential privacy (DP)~\cite{DBLP:conf/icalp/Dwork06}, are established methods to protect privacy of DL models and can enable MLaaS without disclosing either the servers model or the clients input data. These mechanisms preserve privacy at different levels, usually at the expense of accuracy or increased runtime~\cite{boulemtafes2020review,DBLP:journal/aire/how2keep}.

\textit{Homomorphic encryption (HE)} allows to operate on encrypted data without knowledge of the secret decryption key. The concept of HE was already introduced in 1978~\cite{Rivest1978}; however, the first HE scheme, which at least in theory, is capable to evaluate an arbitrary circuit on encrypted data, was only introduced in 2009~\cite{DBLP:conf/stoc/Gentry09}. This scheme allowed for the first time to evaluate an arbitrary number of additions and multiplications on encrypted data. Despite deemed impractical, this work led the way to many improvements~\cite{DBLP:conf/crypto/Brakerski12,DBLP:journals/iacr/FanV12,DBLP:conf/innovations/BrakerskiGV12,DBLP:journals/joc/ChillottiGGI20,DBLP:conf/asiacrypt/CheonKKS17}. The CKKS~\cite{DBLP:conf/asiacrypt/CheonKKS17} encryption scheme introduced novel methods to allow HE for real numbers and is, thus, one of the most promising schemes for machine learning applications. While noting that our framework is compatible with any HE scheme, in this paper we focus on using CKKS and its implementation in the SEAL~\cite{sealcrypto} library (version 4.0.0), a fast, open-source, state-of-the-art HE library maintained by Microsoft Research.

\textit{Differential privacy (DP)} defines a well-established quantitative notion of privacy for single data entries within a database~\cite{dwork2008differential}. The main idea is that a differentially private computation's outcome should be as independent as possible from a single data entry (e.g., by adding noise)~\cite{wood2018differential,DBLP:journal/aire/how2keep}. The privacy guarantee is parameterizable, usually denoted by the letter $\epsilon$. Thus, when applied to training a neural network, the resulting weights will be statistically independent (in the privacy parameter) to the training data.

\subsection{Goals and Contribution}
HE preserves privacy of classification tasks, however it adds a severe performance penalty to the computation, minimizing the achieved throughput. Thus, efficiently training on homomorphically encrypted datasets is still an open challenge. Consequently, combining HE with TL to efficiently preclude privacy issues is another unsolved problem. In this work, we extend private MLaaS to TL scenarios by enabling a private knowledge transfer from rich source domain data into a related sparse target domain task.
In addition, we increase the overall performance for the target domain task by minimizing the size of the networks applied to encrypted data.
Our method, dubbed CryptoTL, ensures that all datasets involved in the knowledge transfer, as well as during classification, are protected. Furthermore, CryptoTL is carefully designed to achieve privacy while having all training and fine-tuning done in plain without HE. We then evaluate our approach in text classification use cases. Since our method only requires HE during a forward-pass of a network, CryptoTL leads to practical fine-tuning and prediction runtimes, confirmed by our extensive benchmarks. Concretely, one forward-pass through the encrypted layers takes roughly $1\,s$ using a notebook CPU. Therefore, we demonstrate for the first time a privacy-preserving TL approach that allows training of accurate DL models on small datasets which are efficient and feasible for real-world use cases despite using HE.

The remainder of this paper is organized as follows.
Section~\ref{sec:relate-work} presents related works at the intersection of privacy and deep TL.
Section~\ref{sec:methodology} describes our methodology.
Section~\ref{sec:experiments} outlines the experimental evaluation of CryptoTL for different scenarios. Section~\ref{sec:results} discusses the results and main findings. Finally, Section~\ref{sec:conclusion} summarizes our contributions.

\section{Related Work}\label{sec:relate-work}

Our method is built at the intersection of TL, HE, and DP.
We start this section with an overview of recent developments of TL and continue with a review of state-of-the-art privacy-preserving machine learning methods.

\paragraph{Transfer learning.}
TL~\cite{pan2009survey} has become a popular research objective in machine learning with wide spread application possibilities.
A broad categorization of the different TL approaches can be found in~\cite{zhuang2020comprehensive}.
Difficulties for TL arise in scenarios where feature spaces of both source and target domains do not overlap in terms of size and domain, so the knowledge transfer cannot be accurately performed~\cite{ben2010theory}.
Further issues include the protection of private source domain data during the knowledge transfer~\cite{zhuang2020comprehensive}.
Nevertheless, TL can be easily adapted for preserving private information, especially in the NLP domain~\cite{DBLP:journal/aire/how2keep}.
CryptoTL, therefore, introduces a TL approach that prevents privacy issues by integrating HE and DP into the neural network model.
Simultaneously, it deals with the heterogeneous feature space problem by using data representations extracted from a state-of-the-art model for language encoding.
\paragraph{Privacy-preserving machine learning.} In recent years, the research increasingly focused on applying privacy-preserving cryptographic protocols and primitives to combat privacy issues during the training and classification of machine learning models. More specifically, secure multiparty computation (MPC) has been applied to both training~\cite{DBLP:conf/sp/MohasselZ17,DBLP:journals/popets/WaghGC19,DBLP:conf/ccs/MohasselR18,DBLP:journals/popets/WaghTBKMR21} and classification~\cite{DBLP:conf/ccs/LiuJLA17,DBLP:conf/sp/Damgard0FKSV19,DBLP:conf/sp/0001RCGR020,DBLP:conf/uss/MishraLSZP20,DBLP:conf/uss/LehmkuhlMSP21,277254}, whereas solutions including HE have mostly only successfully been applied to classification tasks~\cite{DBLP:conf/icml/Gilad-BachrachD16,DBLP:conf/uss/JuvekarVC18,DBLP:conf/pldi/DathathriS0LLMM19,DBLP:journals/corr/abs-2101-07078,DBLP:conf/pldi/DathathriKSDLM20,DBLP:conf/icassp/ChouGLGBS20,DBLP:conf/cscml/ChillottiJP21}. Training machine learning models on homomorphically encrypted inputs is still an ongoing research questions. In~\cite{DBLP:conf/cvpr/NandakumarRPH19}, researchers from IBM (including the author of the well-known HElib~\cite{DBLP:journals/iacr/HaleviS20} library for HE) tried to apply HE to the training of a small 3-layer network for classifying the MNIST~\cite{lecun-mnisthandwrittendigit-2010} dataset. They reported runtimes of $1.5$ days for applying one mini-batch with 60 training samples to the network and more then $9$ hours for the same minibatch applied to a minimized version of their network with $98\,\%$ less trainable parameters. As a consequence, the authors conclude that training on encrypted data is still far too slow for real-world applications. Furthermore, in~\cite{DBLP:journals/access/BadawiHMLA20} the authors provide a GPU implementation of the CKKS cryptosystem and use it to speed up training of a DL model for NLP use cases. Despite achieving a classification speedup of factor $22$ compared to just using CPU's, training a model still took them more than 5 days for 2 epochs using 8 GPUs.

In~\cite{DBLP:journals/corr/abs-1811-09953}, the authors apply HE to neural network inference tasks in MLaaS. To minimize the size of the encrypted network they propose to use TL: The client uses public pre-trained neural networks for feature extraction before transmitting the encrypted output features to the server, which then applies fine-tuned layers to this encrypted output features.
Similarly, in~\cite{DBLP:journals/corr/abs-2202-13351}, the authors propose to split the server network in two private parts interleaved by a public part that can be evaluated in plain to speed up classification while using homomorphic encryption. While these approach are somewhat similar to CryptoTL, they can only be used for classification of the client inputs and do not allow a client to fine-tune a network to his own private dataset. In~\cite{DBLP:journals/expert/LiuKXCY20}, the authors describe a framework for private federated TL of DL models based on either HE or MPC. Their approach is vastly different to CryptoTL since their approach applies a bidirectional knowledge transfer during the training of two networks. Their HE based approach only relies on Paillier's~\cite{DBLP:conf/eurocrypt/Paillier99} additive HE scheme and both involved parties repeatedly use the other party to decrypt intermediate values. However, malicious parties can abuse the other party as decryption oracle to decrypt private representations of the input features. Their MPC based approach heavily relies on the precomputation of beaver triples~\cite{DBLP:conf/crypto/Beaver91a} in an offline phase; however, the authors do not give benchmarks of this more expensive offline phase in their paper.
Finally, Hu and Yang~\cite{hu2020privnet} use DP for TL, but not combined with HE.

\section{Privacy-Preserving Transfer Learning: The Methodology}\label{sec:methodology}
\begin{figure}[ht]
  \centering
  \includegraphics[width=0.8\linewidth]{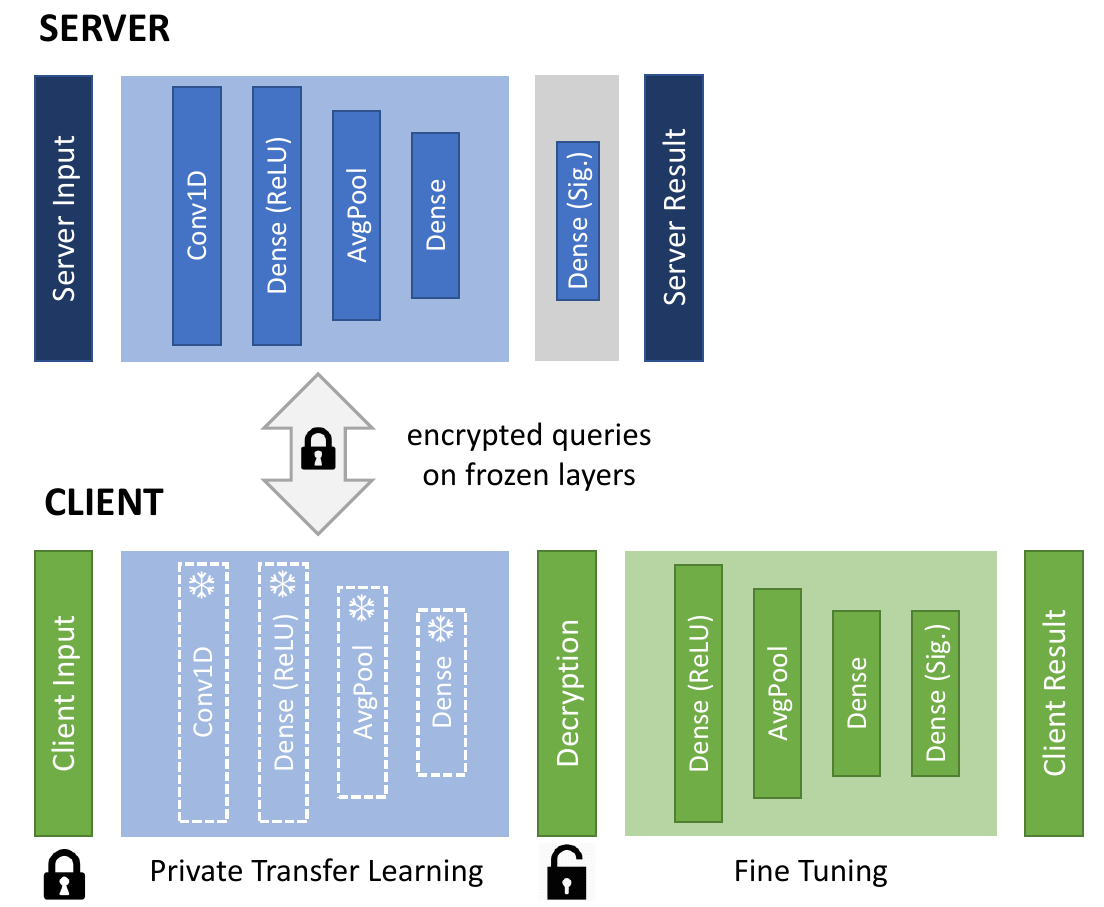}
 \caption{CryptoTL architecture scheme. The lower CNN layers on the server are frozen after training and a client can query homomorphically encrypted input on these layers. The private source domain data can be additionally secured by DP. After the query new top layers are added at the client to fine tune the pre-trained CNN to the target domain data. The fine tuning does not have to be encrypted anymore since it is performed locally at the client.}
 \label{fig:transfer-learning-method}
\end{figure}

\begin{figure}[!htb]
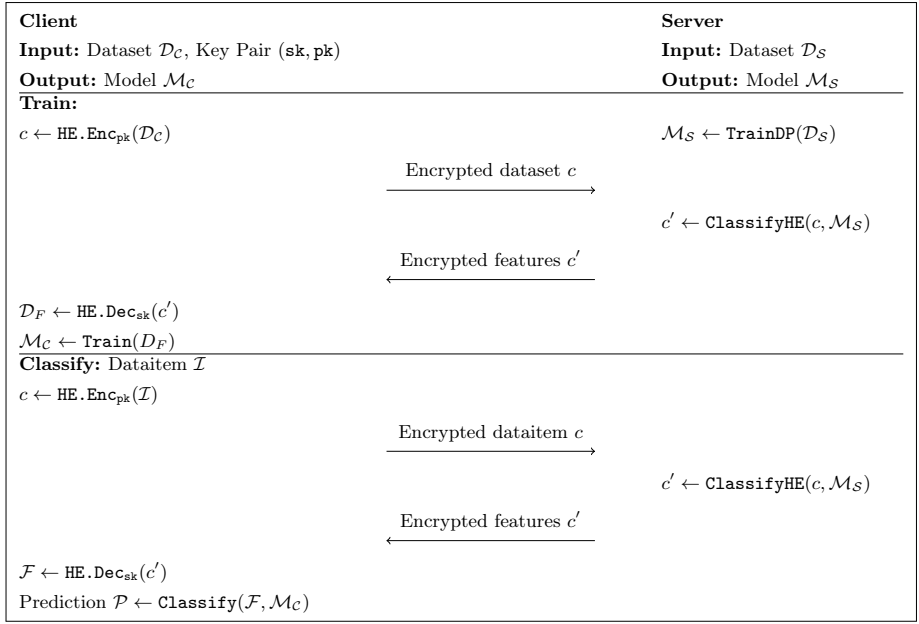

\centering
\resizebox{\linewidth}{!}{%
\fbox{
\pseudocode[colsep=1em,addtolength=5em]{%
    \textbf{Client} \< \< \textbf{Server} \\
    \textbf{Input: } \text{Dataset } \mathcal D_{\mathcal C} \text{, Key Pair } (\texttt{sk}, \texttt{pk})  \< \< \textbf{Input: }  \text{Dataset } \mathcal D_{\mathcal S} \\
    \textbf{Output: } \text{Model } \mathcal M_{\mathcal C}  \< \< \textbf{Output: }  \text{Model } \mathcal M_{\mathcal S} \\
    [][\hline]
    \textbf{Train:} \\
    c \gets \texttt{HE.Enc}_{\texttt{pk}}(\mathcal D_{\mathcal C}) \< \< \mathcal M_{\mathcal S} \gets \texttt{TrainDP}(\mathcal D_{\mathcal S}) \\
    \< \sendmessageright*{\text{Encrypted dataset } c} \\
    \< \< c' \gets \texttt{ClassifyHE}(c, \mathcal M_{\mathcal S}) \\
    \< \sendmessageleft*{\text{Encrypted features } c'} \\
    \mathcal D_F \gets \texttt{HE.Dec}_{\texttt{sk}}(c') \\
    \mathcal M_{\mathcal C} \gets \texttt{Train}(D_F) \\
    [][\hline]
    \textbf{Classify: } \text{Dataitem } \mathcal I \\
    c \gets \texttt{HE.Enc}_{\texttt{pk}}(\mathcal I) \\
    \< \sendmessageright*{\text{Encrypted dataitem } c} \\
    \< \< c' \gets \texttt{ClassifyHE}(c, \mathcal M_{\mathcal S}) \\
    \< \sendmessageleft*{\text{Encrypted features } c'} \\
    \mathcal F \gets \texttt{HE.Dec}_{\texttt{sk}}(c') \\
    \text{Prediction }\mathcal P \gets \texttt{Classify}(\mathcal F, \mathcal M_{\mathcal C})
}
}
}
\caption{CryptoTL protocol to train a model and classify data.\label{alg:cryptotl}}
\end{figure}

\Cref{fig:transfer-learning-method} and \Cref{alg:cryptotl} depict the CryptoTL framework, which is outlined as follows.
In our scenario, a server stores the source dataset privately, whereas the target dataset is held by a client.
Both datasets potentially involve sensitive data demanding protection.
The server owns a private model trained on his private source dataset that serves as a basis for the knowledge transfer. This model is trained by adding DP noise to the gradients during training using differentially private stochastic gradient descent (DPSGD)~\cite{abadi2016deep} optimization. This DP protects against leaking the servers dataset when the model is queried.
The client's network then consists of (part of) the server's network, plus some additional layers.
To protect the privacy of the source model, the server does not disclose the network to the client but allows him to query output features for the clients input.
The client does not want to disclose their dataset and sends the queries encrypted using a HE scheme. The server then applies the network on the encrypted input and produces encrypted output features, which can only be decrypted by the client.

During fine-tuning, the client freezes the layers of the server and training of the top layers is done in plain based on the servers output features. Consequently, with our method, training is always done in plain and does not suffer from the performance penalties from the HE scheme. Further, when classifying a new input, the client first queries the server network with HE (similar to private MLaaS) before applying the top layers in plain.
Therefore, using our method does not require that the whole CNN is applied on encrypted data, but only the server's part. This decreases classification latency compared to applying the full network to encrypted data.

To summarize, the security model of CryptoTL is that the client does not leak any information on its dataset to the server due to the usage of HE. In other words, even a malicious server who arbitrarily deviates from the protocol does not learn anything about the client data. The server dataset is protected from the client by leveraging differential privacy during the training of the server model, hence even a malicious client is not able to reconstruct the server dataset. Thus, CryptoTL achieves privacy against both, a malicious server and a malicious client.

\subsection{Encrypting a Convolutional Neural Network}\label{sub:encryption}
Our CryptoTL framework allows the training of a model based on a sparse dataset using transfer learning while keeping all involved data private. HE is, thereby, applied as black box to a given model architecture. Nonetheless, we want to show the efficiency of CryptoTL by giving a specific instantiation using the CKKS homomorphic encryption scheme while noting that CryptoTL is compatible with other HE schemes and model architectures as well.

\paragraph{The CKKS HE scheme.}
The CKKS cryptosystem is a (Ring-) Learning With Errors (LWE~\cite{DBLP:conf/stoc/Regev05}, RLWE~\cite{DBLP:conf/eurocrypt/LyubashevskyPR10}) based HE scheme which introduced novel methods to allow HE for real numbers. The main idea behind CKKS is to first scale the input real numbers by a scale $\Delta$, round them to the next integer, and encode the noise, which is required for security in (R)LWE based encryption schemes, in the least significant digits of the encoded real numbers. A rescaling operation applied after a multiplication then is equivalent to rounding the result, limiting the noise growth in the ciphertext and getting rid of unnecessary least significant digits. Rescaling, however, can only be performed a limited number of times with this number depending on the used CKKS parameters\footnote{The lattice dimension $N$ and the modulus $q$. Larger $q$ allow more rescaling operations and/or a larger $\Delta$, while decreasing security. Bigger $N$ imply more security and thus allows larger $q$, but increase runtime.}. As a result, the number of consecutive multiplications is also limited by these parameters. Increasing them to support more rescaling operations (and consequently multiplications) in general results in a significantly increased runtime. Therefore, the multiplicative depth of the circuit (in our case the machine learning model) is the main performance metric in CKKS.
While the novel methods in CKKS allow HE over real numbers, they also introduce approximation errors into the computation. In general, the larger the scale $\Delta$ the smaller the impact of the approximation errors.\footnote{\label{foot:ckks_acc} In SEAL, $\Delta$ can not arbitrarily increased to support arbitrary accuracy. $\Delta$, thereby, is limited to $60\,$bit, but one also has to consider reserving enough precision ($60-\text{log}_2\Delta$ bits) before the decimal point.} However, increasing $\Delta$ also requires larger CKKS parameters to achieve the same security, further increasing runtime. Therefore, using CKKS in machine learning applications requires choosing a trade-off between runtime, accuracy of computations, and security.

\paragraph{SIMD Encoding.}
In the CKKS encryption scheme, one can encrypt a vector of real numbers into only one ciphertext. Homomorphic additions and multiplications then affect these encrypted vectors elementwise, similar to SIMD instructions on modern CPUs. Additionally, one can also perform a cyclic rotation of the encrypted vectors, allowing to implement efficient SIMD algorithms for fast classifications. HE adds a lot of computational overhead to the classification task. Therefore, one should use this SIMD capabilities of the CKKS encryption scheme to speed up homomorphic classifications. We use the following SIMD-algorithms in our experimental evaluation of the CryptoTL framework, while noting that CryptoTL can be applied to different use cases requiring additional layer implementations, such as 2D convolutions.

\paragraph{Dense-Layer.} We use the babystep-giantstep optimized diagonal method~\cite{DBLP:conf/crypto/HaleviS14,DBLP:conf/eurocrypt/HaleviS15,DBLP:conf/crypto/HaleviS18} to efficiently implement the matrix-vector product of a plain matrix $M\in\R^{t\times t}$ and an encrypted vector $\vec{x}\in\R^t$:
\begin{equation}
  M\vec{x} = \sum_{k=0}^{t_2-1} \texttt{rot}_{(kt_1)}\left(\sum_{j=0}^{t_1-1}\texttt{diag}'_{(kt_1 + j)}(M)\circ \texttt{rot}_j(\vec{x})\right), \label{eq:bsgs}
\end{equation}
where $t = t_1\cdot t_2$, $\texttt{rot}_j(\vec{x})$ rotates the vector $\vec{x}$ by $j$ steps to the left, $\texttt{diag}'_i(M) = \texttt{rot}_{(-\floor{i/t_1}\cdot t_1)}\left(\texttt{diag}_i(M)\right)$\ifdefined\full\else{}, and $\texttt{diag}_i(M)$ expresses the $i$-th diagonal of a matrix $M$ in a vector of size $t$, with $i=0$ being the main diagonal. As a result, a matrix multiplication requires $t_1 + t_2 - 2$ rotations, $t$ homomorphic plaintext-ciphertext multiplications, and $t-1$ additions, and the total depth is 1 plaintext-ciphertext multiplication. We pad non-square matrices with zeros in our implementations.

\paragraph{1D Convolutions.} We implement a 1D variant of the packed-SISO Convolution from~\cite{DBLP:conf/uss/JuvekarVC18}. As a result, the convolutional layer with filter size $f$ requires $f$ homomorphic plaintext-ciphertext multiplications, $f - 1$ rotations, $f$ additions and the total depth is 1 plaintext-ciphertext multiplication.

\paragraph{Average Pool.} We use pooling layers with a stride $s=1$ and size $f$, which can be efficiently implemented with $f -1$ homomorphic rotations, $f$ additions, and 1 plaintext-ciphertext multiplication:
$
    \vec{x}_o = \frac{1}{f} \cdot \sum_{i=0}^f \texttt{rot}_{-i}(\vec{x})
$.

\paragraph{ReLU.}
By leveraging the SIMD-capabilities of CKKS the ReLU activation can be applied to all neurons in parallel. However, CKKS only allows homomorphic additions, multiplications, and rotations and thus requires the approximation of any complex function in the encrypted layers in a basis of polynomials.
Taylor or Chebychev expansions are typically used for that, the latter sometimes giving slightly better results with faster convergence.
In our case it was sufficient to only approximate the ReLU activation function and its derivative for the gradients.
We have tested different approximations and found that already a low number of polynomials gives good results, as long as divergences at the boundaries of the approximation intervals can be excluded.
We have used the Matlab toolbox Chebfun~\cite{driscoll2014chebfun} for an expansion of the ReLU activation into a monomial basis up to terms of degree 3, namely
\begin{equation}
    \text{ReLU}_\text{approx}(z) = -0.0061728z^3 + 0.092593z^2 + 0.59259z + 0.49383
\end{equation}

\paragraph{Multiple Predictions.}
The algorithms discussed so far only require a subset of the available SIMD slots. Consequently, we can use the remaining SIMD slots to classify multiple data items in parallel. In our concrete network, the dense layer requires the most SIMD slots (twice the number of neurons), so we can classify $p = \left\lfloor \frac{\texttt{\#slots}}{2 \cdot \texttt{\#neurons}} \right\rfloor$ items in parallel, where \texttt{\#neurons} is the number of neurons of the largest dense layer.

\paragraph{Multiplicative Depth.} The total multiplicative depth (including plaintext-ciphertext multiplications) of homomorphically evaluating the frozen network from \Cref{fig:transfer-learning-method} is 6. To support this depth (i.e., 6 rescaling operations) in SEAL, the ciphertext modulus $q$ must be split into $8$ primes $q_i$, where 6 primes are $q_i\approx \Delta$ and the remaining two primes are of size $\gamma - \text{log}_2\Delta$ bits, with $\gamma$ defining the pre-comma precision of the result.

\subsection{Adding Differential Privacy to Model Gradients}\label{sub:add-dp}
To grant the server model additional privacy guarantees, we add DP noise to the gradient of CryptoTL during the model optimization.
We follow the DPSGD~\cite{abadi2016deep} optimization approach, which limits the influence of the training data for training a model with parameters $\theta$.
The DPSGD optimizer computes the gradient$\nabla\mathcal{L}(\theta,\vec{v})$ for a subset of randomly selected inputs, clips each gradient's $l_2$ norm, adds DP noise, and takes a step in the opposite direction of the perturbed gradient.
When the model converges, the privacy loss of DPSGD is computed.
Reconstruction of the servers original training set is, thereby, hindered by several different considerations.
On the one hand, the weights of the server network are unknown to the client.
On the other hand, the client only gets noisy output features due to the usage of the CKKS cryptosystem, hardening the reconstruction of weights or training data from the output features.
Finally, the server network is trained with DPSGD limits the statistical dependence of server weights to the original training data.

\section{Experiments}\label{sec:experiments}
To test the usefulness of our privacy-preserving TL approach we conduct a broad experimental evaluation across datasets for text classification use cases consisting of binary classification problems, i.e., sentiment analysis and spam detection.\footnote{The source code can be found in \Cref{app:source}.}
Text data involves privacy risks related to copyright and pieces of private information, such as person names, demographic attributes, and location, which might be sensitive data and should be protected.
In this section, we describe the settings of the experimental evaluation for CryptoTL. We want to note that CryptoTL, however, has the potential to be applied to additional ML tasks which profit from privacy-preserving transfer learning, such as, e.g., image classification and credit-risk assessment.

\subsection{Experimental Setup}

We employ CryptoTL for natural language processing (NLP) text classification tasks and use a state-of-the-art language model for extracting representations as input for the server CNN.
For this task we use a  server CNN which is composed of five layers\footnote{During training of this server network we also add a Dropout layer with rate of 0.2 for better training results.} (see \Cref{fig:transfer-learning-method}), which are trained on the private source domain data, comprising a Convolutional 1-dimensional layer with filter size $f=9$, a Dense layer with approximated ReLU activation and 768 units, an AveragePooling layer of size 3, a Dense layer with 766 units, and an output Dense layer with Sigmoid activation and 2 units.
The first four lower layers of this CNN architecture are used as the basis for the client network and are frozen during training in order to allow knowledge transfer across domains. These layers are queried with homomorphically encrypted inputs from the client to protect its dataset. The fine-tuning on the client side is performed on a set of stacked layers which are trained on the target domain datasets.
In our concrete example these stacked layers are: A second Dense layer with ReLU activation (not approximated, due to evaluation in plain) and 766 units, a MaxPooling layer, a Dropout layer with rate of 0.2, a Dense layer with 764 units, and an output Dense layer with Sigmoid activation and 2 units for prediction.
At the end of the fine-tuning step, the client CNN consists of 8 layers (excluding Dropout layers) as depicted in Fig.~\ref{fig:transfer-learning-method}.

We use Sentence-BERT~\cite{reimers-2019-sentence-bert}, a pre-trained sentence encoder model to extract representations for the NLP datasets which we use in our evaluation (see Table~\ref{tab:summary-of-datasets}).
Sentence-BERT is a modification of BERT~\cite{devlin2019bert} that yields semantically rich representations of sentences by adding a pooling operation to the output of BERT, and fine-tuning it with siamese and triplet networks~\cite{reimers-2019-sentence-bert}.
To hinder the issue of information leakage through BERT~\cite{lehman2021does,song2020information}, we extracted the representations for the datasets using model instances of Sentence-BERT for the training and test sets separately.
At the end of the pre-processing step, each text instance on the NLP datasets was encoded as a 768-dimensional vector by Sentence-BERT. For more details on the pre-processing step we refer to \Cref{app:datasets}.

\subsection{Datasets}\label{sub:datasets}

We conduct the experimental evaluation of our framework on 3 different NLP datasets for text classification tasks of sentiment analysis and spam detection.\footnote{All datasets are publicly available, see \Cref{app:datasets}}
A summary of these datasets can be seen in Table~\ref{tab:summary-of-datasets}.
All of these datasets consist of binary classification problems where the label feature space presents the sentiment polarity associated with a movie review or a tweet, or the presence of spam content in YouTube comments, which can be either positive or negative, true or false.

\begin{table}[ht]
\caption{Overview of the NLP datasets for sentiment analysis and spam detection.}
    \label{tab:summary-of-datasets}
    \centering
    \begin{tabular}{lcc}
    \toprule
       \textbf{Dataset} & \begin{tabular}{c} \textbf{Label} \\ \textbf{Feature} \end{tabular} & \begin{tabular}{c} \textbf{Number of Classes} \slash \\ {[\textbf{Distribution}]} \end{tabular}\\
     \midrule
    IMDB Movie Reviews~\cite{maas-imdb-data} & Sentiment polarity & 2 [24,753/25,247] \\
    Twitter~\cite{dong2014adaptive} (subset) & Sentiment polarity & 2 [1,734/1,733] \\
    YouTube~\cite{alberto2015tubespam} & Spam in comments & 2 [1,005/951] \\
    \bottomrule
    \end{tabular}
\end{table}

\section{Results and Discussion}\label{sec:results}

\subsection{Runtime and Data Communication}
One big advantage of CryptoTL is that during the actual training and fine-tuning of a network no HE is involved, which is why no encrypted back propagation is required. This significantly reduces CryptoTL's training runtime. For fine-tuning, the client needs to query the server once for each (encrypted) element of his target-dataset; afterwards, he can proceed by fine-tuning the upper layers based on the servers output in plain. The total (single-threaded) runtime of the fine-tuning, therefore, is
$
    t_\text{train} = \left\lceil \frac{|\text{target-dataset}|}{p} \right\rceil \cdot t_S + t_{\text{finetune}}
$,
where $t_S$ is the time for one forward pass through the frozen server layers evaluated on encrypted inputs, $t_\text{finetune}$ is the actual fine-tuning time (wihtout encryption) for the upper clientside layers, and $p$ is the number of possible parallel evaluations using the SIMD slots of CKKS. For prediction, the client first needs to query the frozen layers on the server once (using encryption), before applying the fine-tuned upper layers in plain. Thus, the total runtime is $t_\text{predict} = t_S + t_C \approx t_S$.

We depict the (single-threaded) runtime\footnote{\label{foot:pc}We ran the benchmark on a notebook with an Intel i7-1165G7 CPU (2.8\,GHz, turboboost up to 4.7\,GHz) and 32\,GB RAM using gcc version 11.2. Each benchmark used only one thread.} $t_S$ of one forward pass through the frozen serverside layers on homomorphically encrypted input features in \Cref{tab:fhe_cnn}. We, thereby, depict the runtime for two different CKKS parameter sets (depicted by the lattice dimension $N$ which has to be a power-of-two in SEAL, by the modulus $q$, and by the scaling factor $\Delta$) both providing at least $128$\,bit of computational security.\footnote{As seen here \url{http://homomorphicencryption.org/}, verified using \url{https://github.com/malb/lattice-estimator}~\cite{DBLP:journals/jmc/AlbrechtPS15}, and directly enforced by SEAL.} The first parameter set CryptoTL$_{p1}$ thereby uses smaller scaling factors optimized for faster runtime by reducing the accuracy of the scheme, the second parameter set CryptoTL$_{p2}$ trades runtime with a higher CKKS-accuracy. Further, \Cref{tab:fhe_cnn} depicts the runtime for using SEAL configured to use the Intel HEXL library~\cite{IntelHEXL}  with support for the \texttt{AVX512-IFMA52} instruction set of the underlying CPU, and for using SEAL without HEXL and \texttt{AVX512} (dubbed portable mode).

\begin{table}[ht]
\centering
\caption{Single-threaded runtime $t_S$ in seconds, client-to-server communication ($\mathcal{C} \rightarrow \mathcal{S}$) in kB, and server-to-client communication ($\mathcal{S} \rightarrow \mathcal{C}$) in kB for one forward pass through the frozen layers on homomorphically encrypted features for two different CKKS parameter sets providing $128\,$bit security. $p$ are the number of predictions that can be performed in parallel during one forward pass.}
\label{tab:fhe_cnn}
    \addtolength{\tabcolsep}{2pt}
    \begin{tabular}{lcccc|cc|rr}
    \toprule
    \textbf{Parameters} & $\mathbf{N}$ & \textbf{log}$_\mathbf{2}\mathbf{q}$ & \textbf{log}$_\mathbf{2}{\Delta}$ & $\mathbf{p}$ & \multicolumn{2}{c|}{$\mathbf{t_S}$ $\mathbf{[s]}$} & $\mathbf{\mathcal{C} \rightarrow \mathcal{S}}$ & $\mathbf{\mathcal{S} \rightarrow \mathcal{C}}$  \\
    & & & & & \textbf{Portable} & \textbf{\texttt{AVX-512}} & \multicolumn{1}{c}{\textbf{kB}} & \multicolumn{1}{c}{\textbf{kB}} \\
    \midrule
    CryptoTL$_{p1}$ & 8192 & 218 & 25 & 2 & 2.72 & \textbf{1.08} & 235.3 & 82.4 \\
    CryptoTL$_{p2}$ & 16384 & 420 & 50 & 5 & 5.93 & 2.56 & 838.1 & 262.3 \\
    \bottomrule
    \end{tabular}
\end{table}

As mentioned in \Cref{sec:HE}, the multiplicative depth, together with the desired CKKS-accuracy, defines the used HE parameters, with larger parameters significantly increasing the runtime. Once the parameters are fixed, the runtime scales with the homomorphic operations, with ciphertext-ciphertext multiplications, directly followed by homomorphic rotations and plaintext-ciphertext multiplications, being the most expensive operations in CKKS (in terms of runtime). \Cref{tab:fhe_cnn} shows, that the larger parameter set requires more than twice the runtime of the smaller one. We, however, observed in our experiments, that the smaller, more inaccurate, parameter set already produces the same classification accuracy as the network evaluated without HE. Hence, the smaller parameter set is already sufficient for our use case allowing us to use the runtime advantage of the smaller parameter set. The runtime of one homomorphic classification is very fast, despite performing our benchmarks on a notebook CPU. One forward pass through the encrypted network takes $2.72\,s$, which gets reduced to $1.08\,s$ when using the speedup of the \text{AVX-512} instruction set. We want to note, that the runtime could be further reduced by using a GPU implementation of CKKS, such as the one introduced in~\cite{DBLP:journals/access/BadawiHMLA20}.
Overall, the small size of the frozen server layers, as well as the general structure of CryptoTL allows \textit{for the first time} to realize the privacy-preserving training of a new neural network based on a pre-trained private network in feasible runtime, despite using HE.

For completeness we also depict the number of bytes communicated between the client and the server for one forward pass through the CryptoTL network in \Cref{tab:fhe_cnn}. In general, the client needs to transmit a ciphertext of size log$_2(q)\cdot N$\,bits to the server, who then responds with a ciphertext of size log$_2(q_0)\cdot 2\cdot N$\,bits, where $q_0$ is the remaining ciphertext modulus after all rescaling operations. The numbers in \Cref{tab:fhe_cnn} were measured by serializing the corresponding ciphertexts using SEAL, which also includes further compression. The total communication for training using the CryptoTL$_{p1}$ parameters, therefore, is $\left\lceil \frac{|\text{target-dataset}|}{p} \right\rceil \cdot 235.3$\,kB send by the client and $\left\lceil \frac{|\text{target-dataset}|}{p} \right\rceil \cdot 82.4$\,kB responded by the server. For classifcation using CryptoTL$_{p1}$ the communication is only $235.3$\,kB and $82.4$\,kB respectively. This shows that communication between client and server is small enough for practical applications, especially for use cases where the target dataset is small.

\subsection{Transfer Learning Accuracy}\label{sub:tl-accuracy}
In Table~\ref{tab:transfer-learning-results-nlp}, we present the results of applying CryptoTL to sentiment analysis and spam detection datasets.
For all experiments reported in the table, we have firstly set the number of epochs $h$ to 3, run each experiment 16 times\footnote{\label{foot:pc2} We ran the experiments on an Ubuntu 20.04.2 LTS server with 2 x Intel Xeon CPU E5-2630 v4 @ 2.20GHz (20 cores = 40 threads) and 256 GB RAM. Each experiment used one thread.}, and computed the average accuracy at the end of 16 executions.
Secondly, we have used a 10-fold cross-validation to evaluate all models.
Further, we used early stopping with patience of 1 epoch to reduce the likelihood of overfitting on the models.
Finally, we have considered three baselines to compare the results of CryptoTL in the table.

Each column in \Cref{tab:transfer-learning-results-nlp} contains the accuracy of using the dataset in the column heading (i.e., IMDB, YouTube, and Twitter) as the target dataset for TL.
For the CryptoTL tests, each row corresponds to using a different dataset as source dataset (i.e., $\mathcal{S}$: IMDB refers to using IMDB as the source dataset).
The first baseline to compare against our results consists of the PrivFT~\cite{DBLP:journals/access/BadawiHMLA20} model, which performs classification over encrypted data from the IMDB dataset.
In addition, we also compared the CryptoTL results against two CNN models solely trained on the target dataset without TL.
The first CNN model has the same number of layers as CryptoTL, whereas the second solely consists of the upper CryptoTL layers (i.e., the client side layers).
These benchmarks can be found in the last two rows of \Cref{tab:transfer-learning-results-nlp}.
Last, we do not compare CryptoTL with the two different CKKS parameters (\Cref{tab:fhe_cnn}) to a TL network without encryption since in our experiments all these networks produced the same accuracy.

\begin{table}[!ht]
\caption{Transfer learning accuracy for each NLP dataset.}
    \label{tab:transfer-learning-results-nlp}
    \centering
    \addtolength{\tabcolsep}{2pt}
    \begin{tabular}{lcccccc}
    \toprule
     \textbf{Model} & \textbf{IMDB} & \textbf{YouTube} & \textbf{Twitter} \\
    \midrule
     PrivFT~\cite{DBLP:journals/access/BadawiHMLA20} & 89.88 & -- & -- \\
    \midrule
     CryptoTL ($\mathcal{S}$: IMDB) & 86.29 $\pm$ 0.332 & 91.41 $\pm$ 0.575 & 82.65 $\pm$ 1.817 \\
     CryptoTL ($\mathcal{S}$: YouTube) & 80.48 $\pm$ 1.043 & 96.67 $\pm$ 0.786 & 78.03 $\pm$ 2.197 \\
     CryptoTL ($\mathcal{S}$: Twitter) & 83.82 $\pm$ 0.578 & 91.97 $\pm$ 0.994 & 86.70 $\pm$ 0.774\\
     \midrule
     CNN (full, $h$ = 3) & 85.83 $\pm$ 0.904 & 93.30 $\pm$ 0.885 & 87.28 $\pm$ 1.337 \\
     CNN (upper layers, $h$ = 3) & 81.26 $\pm$ 0.320 & 93.96 $\pm$ 0.435 & 82.65 $\pm$ 0.389 \\
    \bottomrule
    \end{tabular}
\end{table}

The results in \Cref{tab:transfer-learning-results-nlp} show positive gains in accuracy over the upper CNN layers baseline when IMDB or Twitter was used as the source dataset for TL.
In these experiments, accuracies on the target datasets rose by, at least, 2.178\% on IMDB and 1.451\% on Twitter.
On the YouTube dataset, CryptoTL achieved accuracies above 90\%, although this dataset hardens the knowledge transfer since it is designed for a different NLP task.
However, when TL was performed using a subset of YouTube as the source dataset, the scores of CryptoTL outperformed all the baselines computed on this dataset.
Additionally, the full CNN without TL and PrivFT demonstrated to be still challenging baselines for TL. The reason for that is, that models benefit from the training and test over the same datasets without shifts in data domain and distribution incurred by using TL. Furthermore, PrivFT is based on the shallow fasttext~\cite{DBLP:conf/eacl/GraveMJB17} architecture and has the advantage of using more input features (i.e., longer sentences).
Nevertheless, CryptoTL achieved competitive results despite our main goal being the achievement of a fast, efficient, and secure transfer learning architecture and \textit{not} advancing the state-of-the-art in classification accuracy.

\begin{figure}
    \centering
    \begin{subfigure}[a]{\textwidth}
         \centering
         \includegraphics[width=\textwidth]{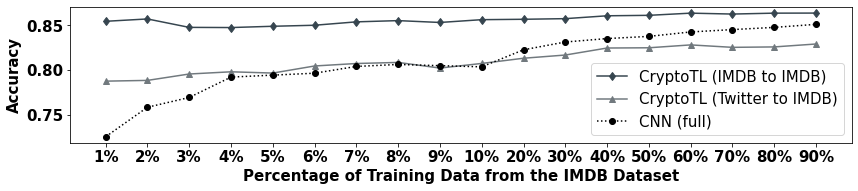}
         \caption{CryptoTL results on the IMDB dataset}
         \label{fig:cryptotl-vs-cnn-imdb}
     \end{subfigure}
     \hfill
      \centering
     \begin{subfigure}[b]{\textwidth}
         \centering
         \includegraphics[width=\textwidth]{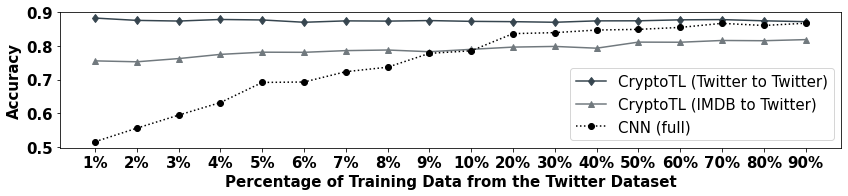}
        \caption{CryptoTL results on the Twitter dataset}
        \label{fig:cryptotl-vs-cnn-twitter}
    \end{subfigure}
     \hfill
    \caption{CryptoTL results against the full CNN (without TL) baseline varying the percentage of data sampled from the training set of the target data (IMDB and Twitter) for fine-tuning.}
    \label{fig:cryptotl-vs-cnn-all}
\end{figure}

To emphasize the functionality of CryptoTL for scenarios featuring scarce data, we reduced the available target training dataset in multiple steps down to 1\% of its original size.
Then, we compared the results of CryptoTL against those of the full CNN without TL in \Cref{fig:cryptotl-vs-cnn-all}.
Figure~\ref{fig:cryptotl-vs-cnn-imdb} depicts the results obtained by the CNN and two settings of CryptoTL on the IMDB dataset.
The CNN model only outperformed CryptoTL pre-trained on Twitter in experiments which used more than 10\% of the training set of IMDB.
Figure~\ref{fig:cryptotl-vs-cnn-twitter} shows similar results for the Twitter dataset.
One can observe that CryptoTL is capable of compensating the reduction in target data, especially if the target training dataset is very small.
These experiments explored the lower bound in data size for fine-tuning encrypted TL models that nonetheless achieved stable performances.
In addition, the overall success of TL strongly depends on the similarity between the source and target datasets.
The higher the similarity, the larger the overall accuracy, even for low target dataset sizes.
Therefore TL between datasets for different NLP tasks remains promising. For more details on the deviation of the achieved accuracy we refer to \Cref{app:results-on-reduced-data}.

\subsection{Differentially Private Optimization}\label{sub:dp-optimization}
DP grants additional privacy guarantees against malicious attacks that aim to disclose private data used to pre-train CryptoTL.
Table~\ref{tab:DP_cryptotl} presents the results of the experiments in which we integrated the DPSGD optimizer into the server-side layers of CryptoTL.
These experiments used IMDB as the source domain data and 100\% of the training set of Twitter as target domain training data.
The accuracy scores in the table were obtained after predictions by CryptoTL on the test set of Twitter.
In the table, $l_2$ refers to the maximum Euclidean norm of the gradients computed over the mini-batches during optimization; noise multiplier controls the amount of noise added to the gradients; and $\epsilon$ measures how strong the privacy guarantees are.
By increasing the amount of noise added to the gradients of CryptoTL, we notice that $\epsilon$ values approach zero, as expected, suggesting stronger privacy guarantees, at the cost of drops in accuracy.
Nevertheless, small amounts of noise, which do not degrade model accuracy at large extents, already suffice to balance this privacy-utility trade-off.

\begin{table}[ht]
\centering
\caption{Differentially private optimization results in TL from IMDB to Twitter.}
\label{tab:DP_cryptotl}
    \begin{tabular}{lcc|cc}
    \toprule
    \textbf{Parameters} & $\mathbf{l_2}$ & \textbf{Noise multiplier} & $\mathbf{\epsilon}$ & \textbf{Accuracy} \\
    \midrule
    CryptoTL$_{(DPSGD)}$ & 0.75 & 0.25 & 33.600 & 81.79 \\
    CryptoTL$_{(DPSGD)}$ & 0.75 & 0.50 & 2.640 & 80.05 \\
    CryptoTL$_{(DPSGD)}$ & 0.75 & 0.75 & 0.829 & 77.16 \\
    CryptoTL$_{(DPSGD)}$ & 0.75 & 1.00 & 0.413 & 66.18 \\
    \midrule
    CryptoTL &  &  &  & 82.65 \\
    \bottomrule
    \end{tabular}
\end{table}

\section{Conclusion}\label{sec:conclusion}
With CryptoTL we have introduced a novel framework for privacy-preserving TL inspired by MLaaS models.
Our solution enables clients with small and private datasets to leverage DL models provided by a server while fully protecting their data.
Utilizing the concept of frozen layers allows us to decouple HE from the training phase of the neural network, resulting in (for the first time) high training efficiency and comparatively low runtime for encrypted queries.
Our experimental evaluation highlights the efficiency gains of CryptoTL for a concrete example. It shows low runtime and high accuracy of CryptoTL, especially for scenarios with reduced and small training datasets.
Our framework, thereby, is general and has the potential to preserve the privacy of a wide range of DL use cases.

\subsubsection*{Acknowledgments.}
This work was supported by EU's Horizon 2020 project TRUSTS grant agreement \numero 871481, and by the "DDAI" COMET Module within the COMET -- Competence Centers for Excellent Technologies Programme, funded by the Austrian Federal Ministry for Transport, Innovation and Technology (bmvit), the Austrian Federal Ministry for Digital and Economic Affairs (bmdw), the Austrian Research Promotion Agency (FFG), the province of Styria (SFG) and partners from industry and academia. The COMET Programme is managed by FFG.

\bibliographystyle{splncs04}
\bibliography{bibliography}

\begin{thebibliography}{10}
\providecommand{\url}[1]{\texttt{#1}}
\providecommand{\urlprefix}{URL }
\providecommand{\doi}[1]{https://doi.org/#1}

\bibitem{abadi2016deep}
Abadi, M., Chu, A., Goodfellow, I., McMahan, H.B., Mironov, I., Talwar, K.,
  Zhang, L.: Deep learning with differential privacy. In: Proceedings of the
  2016 ACM SIGSAC Conference on Computer and Communications Security. pp.
  308--318 (2016)

\bibitem{alberto2015tubespam}
Alberto, T.C., Lochter, J.V., Almeida, T.A.: Tubespam: Comment spam filtering
  on youtube. In: 2015 IEEE 14th international conference on machine learning
  and applications (ICMLA). pp. 138--143. IEEE (2015)

\bibitem{DBLP:journals/jmc/AlbrechtPS15}
Albrecht, M.R., Player, R., Scott, S.: On the concrete hardness of learning
  with errors. J. Math. Cryptol.  \textbf{9}(3),  169--203 (2015)

\bibitem{DBLP:journals/access/BadawiHMLA20}
Badawi, A.A., Hoang, L., Mun, C.F., Laine, K., Aung, K.M.M.: Privft: Private
  and fast text classification with homomorphic encryption. {IEEE} Access
  \textbf{8},  226544--226556 (2020)

\bibitem{BAIK2020101431}
Baik, J.S.: Data privacy against innovation or against discrimination?: The
  case of the california consumer privacy act (ccpa). Telematics and
  Informatics  \textbf{52},  101431 (2020).
  \doi{https://doi.org/10.1016/j.tele.2020.101431},
  \url{https://www.sciencedirect.com/science/article/pii/S0736585320300903}

\bibitem{DBLP:conf/crypto/Beaver91a}
Beaver, D.: Efficient multiparty protocols using circuit randomization. In:
  {CRYPTO}. Lecture Notes in Computer Science, vol.~576, pp. 420--432. Springer
  (1991)

\bibitem{ben2010theory}
Ben-David, S., Blitzer, J., Crammer, K., Kulesza, A., Pereira, F., Vaughan,
  J.W.: A theory of learning from different domains. Machine learning
  \textbf{79}(1),  151--175 (2010)

\bibitem{IntelHEXL}
Boemer, F., Kim, S., Seifu, G., de~Souza, F.D., Gopal, V., et~al.: {I}ntel
  {HEXL} (release 1.2). \url{https://github.com/intel/hexl} (2021)

\bibitem{boulemtafes2020review}
Boulemtafes, A., Derhab, A., Challal, Y.: A review of privacy-preserving
  techniques for deep learning. Neurocomputing  \textbf{384},  21--45 (2020)

\bibitem{DBLP:conf/crypto/Brakerski12}
Brakerski, Z.: Fully homomorphic encryption without modulus switching from
  classical gapsvp. In: {CRYPTO}. LNCS, vol.~7417, pp. 868--886. Springer
  (2012)

\bibitem{DBLP:conf/innovations/BrakerskiGV12}
Brakerski, Z., Gentry, C., Vaikuntanathan, V.: (leveled) fully homomorphic
  encryption without bootstrapping. In: {ITCS}. pp. 309--325. {ACM} (2012)

\bibitem{277254}
Chandran, N., Gupta, D., Obbattu, S.L.B., Shah, A.: {SIMC}: {ML} inference
  secure against malicious clients at {Semi-Honest} cost. In: {USENIX} Security
  Symposium. pp. 1361--1378. {USENIX} Association (2022)

\bibitem{DBLP:conf/asiacrypt/CheonKKS17}
Cheon, J.H., Kim, A., Kim, M., Song, Y.S.: Homomorphic encryption for
  arithmetic of approximate numbers. In: {ASIACRYPT} {(1)}. LNCS, vol. 10624,
  pp. 409--437. Springer (2017)

\bibitem{DBLP:journals/joc/ChillottiGGI20}
Chillotti, I., Gama, N., Georgieva, M., Izabach{\`{e}}ne, M.: {TFHE:} fast
  fully homomorphic encryption over the torus. J. Cryptol.  \textbf{33}(1),
  34--91 (2020)

\bibitem{DBLP:conf/cscml/ChillottiJP21}
Chillotti, I., Joye, M., Paillier, P.: Programmable bootstrapping enables
  efficient homomorphic inference of deep neural networks. In: {CSCML}. Lecture
  Notes in Computer Science, vol. 12716, pp. 1--19. Springer (2021)

\bibitem{DBLP:journals/corr/abs-1811-09953}
Chou, E., Beal, J., Levy, D., Yeung, S., Haque, A., Fei{-}Fei, L.: Faster
  cryptonets: Leveraging sparsity for real-world encrypted inference. CoRR
  \textbf{abs/1811.09953} (2018)

\bibitem{DBLP:conf/icassp/ChouGLGBS20}
Chou, E.J., Gururajan, A., Laine, K., Goel, N.K., Bertiger, A., Stokes, J.W.:
  Privacy-preserving phishing web page classification via fully homomorphic
  encryption. In: {ICASSP}. pp. 2792--2796. {IEEE} (2020)

\bibitem{EUdataregulations2018}
Commission, E.: 2018 reform of eu data protection rules.
  \url{https://ec.europa.eu/commission/sites/beta-political/files/data-protection-factsheet-changes_en.pdf}
  (2018), date: 2018-05-25, URL Date: 2019-06-17

\bibitem{DBLP:conf/sp/Damgard0FKSV19}
Damg{\aa}rd, I., Escudero, D., Frederiksen, T.K., Keller, M., Scholl, P.,
  Volgushev, N.: New primitives for actively-secure {MPC} over rings with
  applications to private machine learning. In: {IEEE} Symposium on Security
  and Privacy. pp. 1102--1120. {IEEE} (2019)

\bibitem{DBLP:conf/pldi/DathathriKSDLM20}
Dathathri, R., Kostova, B., Saarikivi, O., Dai, W., Laine, K., Musuvathi, M.:
  {EVA:} an encrypted vector arithmetic language and compiler for efficient
  homomorphic computation. In: {PLDI}. pp. 546--561. {ACM} (2020)

\bibitem{DBLP:conf/pldi/DathathriS0LLMM19}
Dathathri, R., Saarikivi, O., Chen, H., Laine, K., Lauter, K.E., Maleki, S.,
  Musuvathi, M., Mytkowicz, T.: {CHET:} an optimizing compiler for
  fully-homomorphic neural-network inferencing. In: {PLDI}. pp. 142--156. {ACM}
  (2019)

\bibitem{devlin2019bert}
Devlin, J., Chang, M.W., Lee, K., Toutanova, K.: Bert: Pre-training of deep
  bidirectional transformers for language understanding. In: Proceedings of the
  2019 Conference of the North American Chapter of the Association for
  Computational Linguistics: Human Language Technologies, Volume 1 (Long and
  Short Papers). pp. 4171--4186 (2019)

\bibitem{dong2014adaptive}
Dong, L., Wei, F., Tan, C., Tang, D., Zhou, M., Xu, K.: Adaptive recursive
  neural network for target-dependent twitter sentiment classification. In: The
  52nd Annual Meeting of the Association for Computational Linguistics (ACL).
  ACL (2014)

\bibitem{driscoll2014chebfun}
Driscoll, T.A., Hale, N., Trefethen, L.N.: Chebfun guide (2014)

\bibitem{DBLP:conf/icalp/Dwork06}
Dwork, C.: Differential privacy. In: {ICALP} {(2)}. Lecture Notes in Computer
  Science, vol.~4052, pp. 1--12. Springer (2006)

\bibitem{dwork2008differential}
Dwork, C.: Differential privacy: A survey of results. In: International
  conference on theory and applications of models of computation. pp. 1--19.
  Springer (2008)

\bibitem{DBLP:journals/iacr/FanV12}
Fan, J., Vercauteren, F.: Somewhat practical fully homomorphic encryption.
  {IACR} Cryptol. ePrint Arch.  \textbf{2012}, ~144 (2012)

\bibitem{DBLP:conf/stoc/Gentry09}
Gentry, C.: Fully homomorphic encryption using ideal lattices. In: {STOC}. pp.
  169--178. {ACM} (2009)

\bibitem{DBLP:conf/icml/Gilad-BachrachD16}
Gilad{-}Bachrach, R., Dowlin, N., Laine, K., Lauter, K.E., Naehrig, M.,
  Wernsing, J.: Cryptonets: Applying neural networks to encrypted data with
  high throughput and accuracy. In: {ICML}. {JMLR} Workshop and Conference
  Proceedings, vol.~48, pp. 201--210. JMLR.org (2016)

\bibitem{DBLP:conf/crypto/HaleviS14}
Halevi, S., Shoup, V.: Algorithms in helib. In: {CRYPTO} {(1)}. LNCS,
  vol.~8616, pp. 554--571. Springer (2014)

\bibitem{DBLP:conf/eurocrypt/HaleviS15}
Halevi, S., Shoup, V.: Bootstrapping for helib. In: {EUROCRYPT} {(1)}. LNCS,
  vol.~9056, pp. 641--670. Springer (2015)

\bibitem{DBLP:conf/crypto/HaleviS18}
Halevi, S., Shoup, V.: Faster homomorphic linear transformations in helib. In:
  {CRYPTO} {(1)}. LNCS, vol. 10991, pp. 93--120. Springer (2018)

\bibitem{DBLP:journals/iacr/HaleviS20}
Halevi, S., Shoup, V.: Design and implementation of helib: a homomorphic
  encryption library. {IACR} Cryptol. ePrint Arch.  \textbf{2020}, ~1481 (2020)

\bibitem{heetal2009modelinversion}
He, Z., Zhang, T., Lee, R.B.: Model inversion attacks against collaborative
  inference. Proceedings of the 35th Annual Computer Security Applications
  Conference p. 148–162 (2019)

\bibitem{hu2020privnet}
Hu, G., Yang, Q.: Privnet: Safeguarding private attributes in transfer learning
  for recommendation. In: Proceedings of the 2020 Conference on Empirical
  Methods in Natural Language Processing: Findings. pp. 4506--4516 (2020)

\bibitem{DBLP:conf/eacl/GraveMJB17}
Joulin, A., Grave, E., Bojanowski, P., Mikolov, T.: Bag of tricks for efficient
  text classification. In: {EACL} {(2)}. pp. 427--431. Association for
  Computational Linguistics (2017)

\bibitem{DBLP:conf/uss/JuvekarVC18}
Juvekar, C., Vaikuntanathan, V., Chandrakasan, A.: {GAZELLE:} {A} low latency
  framework for secure neural network inference. In: {USENIX} Security
  Symposium. pp. 1651--1669. {USENIX} Association (2018)

\bibitem{DBLP:conf/sp/0001RCGR020}
Kumar, N., Rathee, M., Chandran, N., Gupta, D., Rastogi, A., Sharma, R.:
  Cryptflow: Secure tensorflow inference. In: {IEEE} Symposium on Security and
  Privacy. pp. 336--353. {IEEE} (2020)

\bibitem{lecun-mnisthandwrittendigit-2010}
LeCun, Y., Cortes, C.: {MNIST} handwritten digit database (2010),
  \url{http://yann.lecun.com/exdb/mnist/}

\bibitem{lehman2021does}
Lehman, E., Jain, S., Pichotta, K., Goldberg, Y., Wallace, B.C.: Does bert
  pretrained on clinical notes reveal sensitive data? In: Proceedings of the
  2021 Conference of the North American Chapter of the Association for
  Computational Linguistics: Human Language Technologies. pp. 946--959 (2021)

\bibitem{DBLP:conf/uss/LehmkuhlMSP21}
Lehmkuhl, R., Mishra, P., Srinivasan, A., Popa, R.A.: Muse: Secure inference
  resilient to malicious clients. In: {USENIX} Security Symposium. pp.
  2201--2218. {USENIX} Association (2021)

\bibitem{DBLP:conf/ccs/LiuJLA17}
Liu, J., Juuti, M., Lu, Y., Asokan, N.: Oblivious neural network predictions
  via minionn transformations. In: {CCS}. pp. 619--631. {ACM} (2017)

\bibitem{DBLP:journals/expert/LiuKXCY20}
Liu, Y., Kang, Y., Xing, C., Chen, T., Yang, Q.: A secure federated transfer
  learning framework. {IEEE} Intell. Syst.  \textbf{35}(4),  70--82 (2020)

\bibitem{DBLP:conf/eurocrypt/LyubashevskyPR10}
Lyubashevsky, V., Peikert, C., Regev, O.: On ideal lattices and learning with
  errors over rings. In: {EUROCRYPT}. LNCS, vol.~6110, pp. 1--23. Springer
  (2010)

\bibitem{maas-imdb-data}
Maas, A.L., Daly, R.E., Pham, P.T., Huang, D., Ng, A.Y., Potts, C.: Learning
  word vectors for sentiment analysis. In: Proceedings of the 49th Annual
  Meeting of the Association for Computational Linguistics: Human Language
  Technologies. pp. 142--150. Association for Computational Linguistics,
  Portland, Oregon, USA (June 2011)

\bibitem{DBLP:conf/uss/MishraLSZP20}
Mishra, P., Lehmkuhl, R., Srinivasan, A., Zheng, W., Popa, R.A.: Delphi: {A}
  cryptographic inference service for neural networks. In: {USENIX} Security
  Symposium. pp. 2505--2522. {USENIX} Association (2020)

\bibitem{DBLP:conf/ccs/MohasselR18}
Mohassel, P., Rindal, P.: Aby\({}^{\mbox{3}}\): {A} mixed protocol framework
  for machine learning. In: {CCS}. pp. 35--52. {ACM} (2018)

\bibitem{DBLP:conf/sp/MohasselZ17}
Mohassel, P., Zhang, Y.: Secureml: {A} system for scalable privacy-preserving
  machine learning. In: {IEEE} Symposium on Security and Privacy. pp. 19--38.
  {IEEE} Computer Society (2017)

\bibitem{DBLP:conf/cvpr/NandakumarRPH19}
Nandakumar, K., Ratha, N.K., Pankanti, S., Halevi, S.: Towards deep neural
  network training on encrypted data. In: {CVPR} Workshops. pp. 40--48.
  Computer Vision Foundation / {IEEE} (2019)

\bibitem{DBLP:conf/eurocrypt/Paillier99}
Paillier, P.: Public-key cryptosystems based on composite degree residuosity
  classes. In: {EUROCRYPT}. Lecture Notes in Computer Science, vol.~1592, pp.
  223--238. Springer (1999)

\bibitem{pan2009survey}
Pan, S.J., Yang, Q.: A survey on transfer learning. IEEE Transactions on
  knowledge and data engineering  \textbf{22}(10),  1345--1359 (2009)

\bibitem{DBLP:journals/corr/abs-2202-13351}
Pereteanu, G., Alansary, A., Passerat{-}Palmbach, J.: Split {HE:} fast secure
  inference combining split learning and homomorphic encryption. CoRR
  \textbf{abs/2202.13351} (2022)

\bibitem{DBLP:conf/stoc/Regev05}
Regev, O.: On lattices, learning with errors, random linear codes, and
  cryptography. In: {STOC}. pp. 84--93. {ACM} (2005)

\bibitem{reimers-2019-sentence-bert}
Reimers, N., Gurevych, I.: Sentence-bert: Sentence embeddings using siamese
  bert-networks. In: Proceedings of the 2019 Conference on Empirical Methods in
  Natural Language Processing. Association for Computational Linguistics (11
  2019), \url{https://arxiv.org/abs/1908.10084}

\bibitem{Rivest1978}
Rivest, R.L., Adleman, L., Dertouzos, M.L.: On data banks and privacy
  homomorphisms. Foundations of Secure Computation, Academia Press pp. 169--179
  (1978)

\bibitem{sealcrypto}
{M}icrosoft {SEAL} (release 4.0). \url{https://github.com/Microsoft/SEAL} (Mar
  2022), microsoft Research, Redmond, WA.

\bibitem{song2020information}
Song, C., Raghunathan, A.: Information leakage in embedding models. In:
  Proceedings of the 2020 ACM SIGSAC Conference on Computer and Communications
  Security. pp. 377--390 (2020)

\bibitem{DBLP:journal/aire/how2keep}
Sousa, S., Kern, R.: How to keep text private? a systematic review of deep
  learning methods for privacy-preserving natural language processing.
  Artificial Intelligence Review pp. 1--66 (2022)

\bibitem{tan2018survey}
Tan, C., Sun, F., Kong, T., Zhang, W., Yang, C., Liu, C.: A survey on deep
  transfer learning. In: International conference on artificial neural
  networks. pp. 270--279. Springer (2018)

\bibitem{DBLP:journals/corr/abs-2101-07078}
Viand, A., Jattke, P., Hithnawi, A.: Sok: Fully homomorphic encryption
  compilers. CoRR  \textbf{abs/2101.07078} (2021)

\bibitem{DBLP:journals/popets/WaghGC19}
Wagh, S., Gupta, D., Chandran, N.: Securenn: 3-party secure computation for
  neural network training. Proc. Priv. Enhancing Technol.  \textbf{2019}(3),
  26--49 (2019)

\bibitem{DBLP:journals/popets/WaghTBKMR21}
Wagh, S., Tople, S., Benhamouda, F., Kushilevitz, E., Mittal, P., Rabin, T.:
  Falcon: Honest-majority maliciously secure framework for private deep
  learning. Proc. Priv. Enhancing Technol.  \textbf{2021}(1),  188--208 (2021)

\bibitem{wood2018differential}
Wood, A., Altman, M., Bembenek, A., Bun, M., Gaboardi, M., Honaker, J., Nissim,
  K., O'Brien, D.R., Steinke, T., Vadhan, S.: Differential privacy: A primer
  for a non-technical audience. Vand. J. Ent. \& Tech. L.  \textbf{21}, ~209
  (2018)

\bibitem{zhuang2020comprehensive}
Zhuang, F., Qi, Z., Duan, K., Xi, D., Zhu, Y., Zhu, H., Xiong, H., He, Q.: A
  comprehensive survey on transfer learning. Proceedings of the IEEE
  \textbf{109}(1),  43--76 (2020)

\end{thebibliography}

\appendix

\section{Source Code} \label{app:source}
The source code used in our evaluations is publicly available at \url{https://github.com/IAIK/CryptoTL}.

\section{Datasets and Preprocessing}\label{app:datasets}

In this work, we used three NLP datasets for binary classification problems of sentiment analysis and spam detection.
\Cref{tab:summary-of-datasets} presents a summary of these datasets, including their reference, label feature description, and class distribution.
To commence the experimental evaluation, we collected the datasets from public online sources.
First, we obtained the IMDB dataset from the Transformers Python library.\footnote{\url{https://huggingface.co/datasets/imdb}.}
Second, we downloaded the Twitter dataset from the URL provided by~\cite{dong2014adaptive}.\footnote{\url{http://goo.gl/5Enpu7}.}
Finally, we obtained the YouTube dataset from UCI Machine Learning Repository.\footnote{\url{https://archive.ics.uci.edu/ml/datasets/YouTube+Spam+Collection}.}

\paragraph{Dataset pre-processing and Sentence-BERT.}

We used Sentence-BERT to extract the representations for the sentences in the datasets.
In the data pre-processing step, we input each text instance from the datasets to Sentence-BERT following the guidelines in the model documentation.\footnote{\url{https://www.sbert.net/examples/applications/computing-embeddings/README.html}.}
We used the following parameters for Sentence-BERT:
\begin{itemize}
    \item tokenizer: "sentence-transformers/all-distilroberta-v1".
    \item model: "sentence-transformers/all-distilroberta-v1".
    \item max\_length: 512.
\end{itemize}
Finally, we extracted 768-dimensional embeddings for the sentences in the datasets.

\section{Transfer Learning Accuracy for Reduced Target Data}~\label{app:results-on-reduced-data}

In this section, we give more details for the results shown in Figure 2. More specifically, we show the distribution of the achieved accuracy when running the training process 16 times. Each boxplot in \Cref{fig:boxplot-cryptotl-vs-cnn-all}  shows the median accuracy value (the horizontal line in the box), the lower and upper quartiles (lower and upper whiskers), and outliers (points below or above the whiskers). \Cref{fig:boxplot-cryptotl-vs-cnn-all} highlights the advantage of using CryptoTL using related source and target datasets very clearly. For small target datasets, just training a CNN leads to low accuracies, as well as a high deviation in the results. The same behavior can also be observed when more unrelated datasets are combined. The Youtube dataset is used for a different classification task compared to Twitter or IMDB (spam detection vs. sentiment analysis), hence, combining these datasets using CryptoTL also leads to low accuracies and high deviations. However, combining related datasets, e.g., Twitter and IMDB, leads to high accuracies and low deviations, even on small datasets.

\begin{figure}
    \centering
    \begin{subfigure}[a]{\textwidth}
         \centering
         \includegraphics[width=\textwidth]{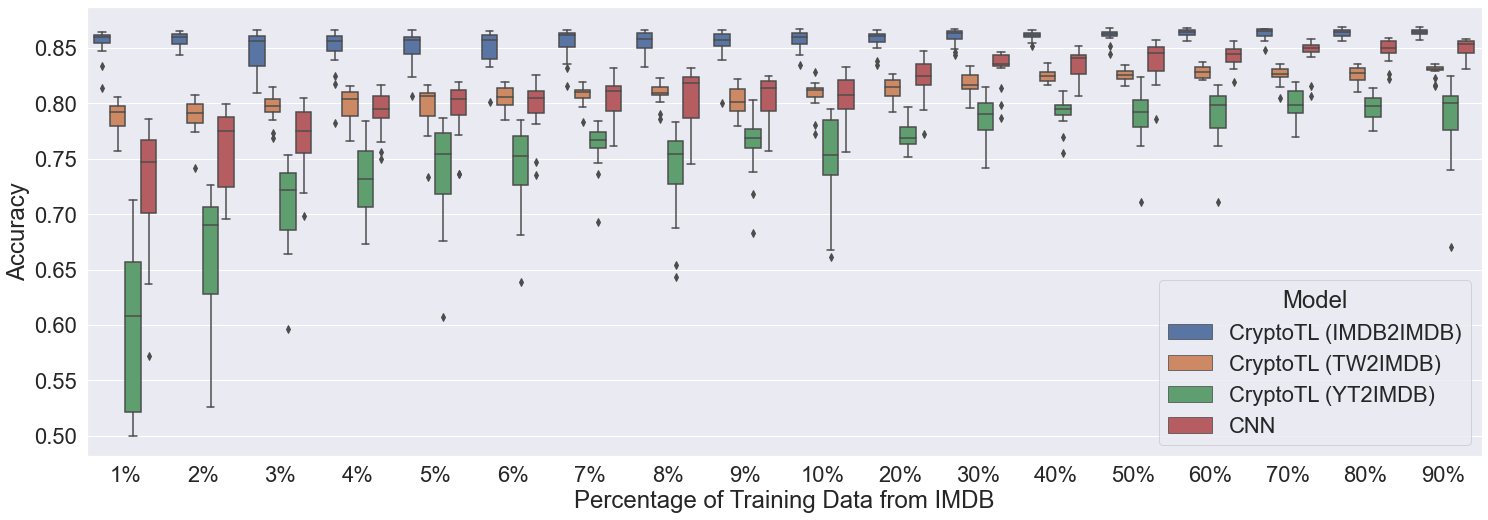}
         \caption{CryptoTL results on the IMDB dataset}
         \label{fig:boxplot-cryptotl-vs-cnn-imdb}
     \end{subfigure}
     \hfill
      \centering
     \begin{subfigure}[b]{\textwidth}
         \centering
         \includegraphics[width=\textwidth]{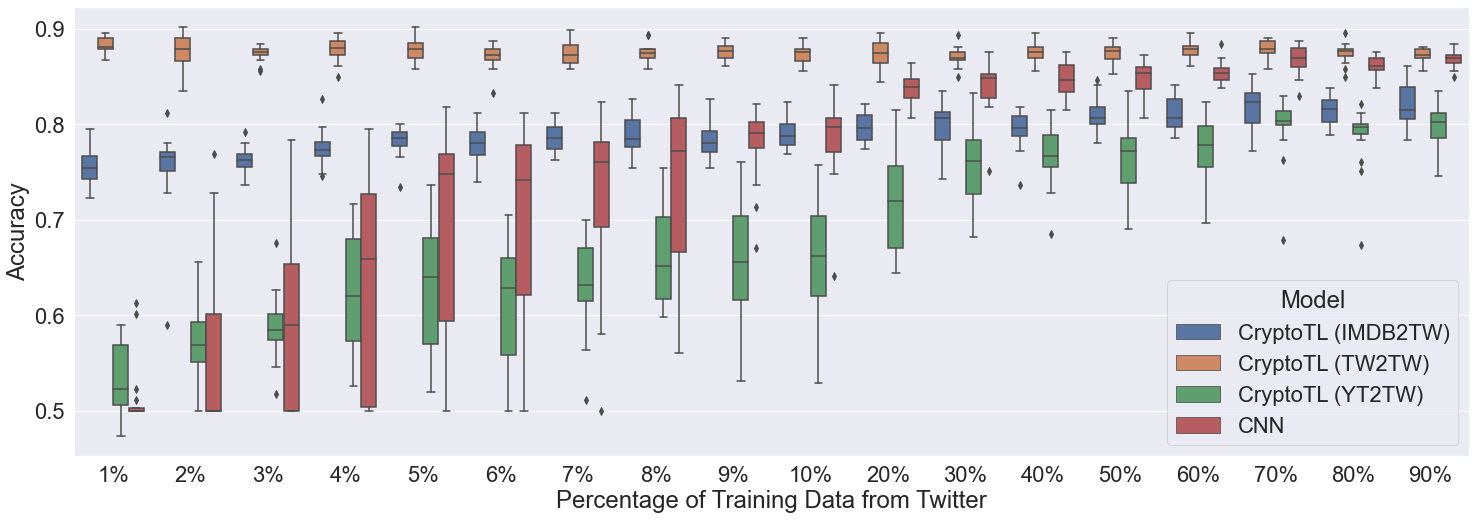}
        \caption{CryptoTL results on the Twitter dataset}
        \label{fig:boxplot-cryptotl-vs-cnn-twitter}
    \end{subfigure}
     \hfill
      \centering
     \begin{subfigure}[c]{\textwidth}
         \centering
         \includegraphics[width=\textwidth]{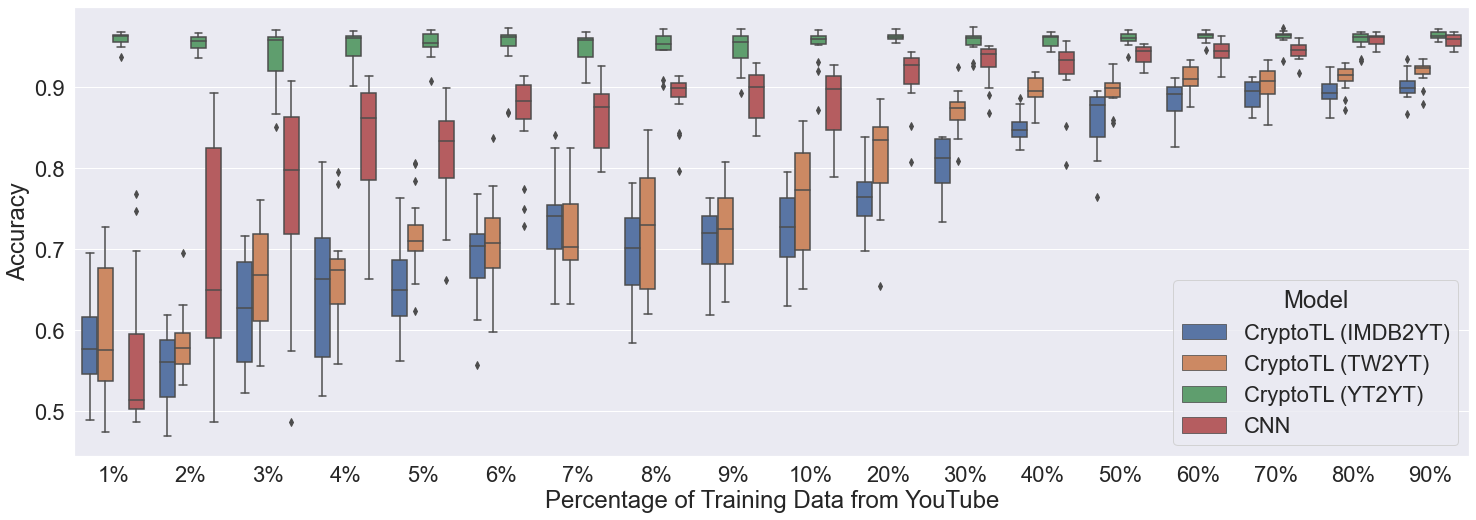}
    \caption{CryptoTL results on the YouTube dataset}
    \label{fig:boxplot-cryptotl-vs-cnn-youtube}
    \end{subfigure}
    \caption{CryptoTL results against the full CNN (without TL) baseline varying the percentage of data sampled from the training set of the target data (IMDB, Twitter, and YouTube) for fine-tuning.}
    \label{fig:boxplot-cryptotl-vs-cnn-all}
\end{figure}


\end{document}